\documentclass[prl,preprint,aps,superscriptaddress,showpacs,floatfix]{revtex4}
\usepackage{graphicx}

\begin{document}

\title{Anomalous proximity effect in gold coated (110) $YBa_2Cu_3O_{7-\delta}$
films: Penetration of the Andreev bound states.}

\author{Itay Asulin}
\affiliation{Racah Institute of Physics, The Hebrew University,
Jerusalem 91904, Israel}

\author{Amos Sharoni}
\affiliation{Racah Institute of Physics, The Hebrew University,
Jerusalem 91904, Israel}

\author{Ofer Yulli}
\affiliation{Racah Institute of Physics, The Hebrew University,
Jerusalem 91904, Israel}

\author{Gad Koren}
\affiliation{Department of Physics, Technion - Israel Institute of
Technology, Haifa 32000, Israel}

\author{Oded Millo}
\email{milode@vms.huji.ac.il} \affiliation{Racah Institute of
Physics, The Hebrew University, Jerusalem 91904, Israel}

\begin{abstract}
Scanning tunneling spectroscopy of (110)
$YBa_2Cu_3O_{7-\delta}/Au$ bi-layers reveal a proximity effect
markedly different from the conventional one. While
proximity-induced mini-gaps rarely appear in the Au layer, the
Andreev bound states clearly penetrate into the metal. Zero bias
conductance peaks are measured on Au layers thinner than 7 nm with
magnitude similar to those detected on the bare superconductor
films. The peaks then decay abruptly with Au thickness and
disappear above 10 nm. This length is shorter than the normal
coherence length and corresponds to the (ballistic) mean free
path.
\end{abstract}


\maketitle

The mutual effect of a superconductor (S) in good electrical
contact with a normal metal (N), a phenomenon known as the
proximity effect (PE), is one of the most intriguing fundamental
properties of superconductors that also yields various
applications \cite{1,2}. The PE has been studied extensively for
conventional (\textit{s}-wave) superconductors in contact with a
normal metal. In such systems, an abrupt change in the pair
potential, from a finite value on the S side to zero on the N
side, leads to a smooth change of the pair amplitude, from its
full bulk value deep in the S side to zero at a distance
characterized by the normal coherence length, $\xi_n$
\cite{2,3,4,5,6}. The pair amplitude is proportional to the gap in
the quasi-particle density of states (DOS) and can thus be
monitored by tunneling spectroscopy.

For an anisotropic \textit{d}-wave superconductor, the
crystallographic orientation of the superconductor surface at the
N-S interface can significantly modify the PE. Sharoni \textit{et
al.} \cite{7}. have shown, using gold coated \textit{c}-axis
$YBa_2Cu_3O_{7-\delta}$ (YBCO) samples, that the PE is primarily
due to the interface between the normal metal and the (100) YBCO
surface, whereas virtually no effect is pertained to the (001)
surface. This facet-selectivity reflects the in plane versus out
of plane anisotropy in the cuprate superconductors, which was also
measured indirectly by other groups \cite{8,9,10}. However, the
proximity-induced order parameter decayed exponentially in a way
typical of conventional N-S proximity systems. In particular, the
measured normal coherence length, $\sim 30$ nm, corresponded to
the dirty limit approximation, $\xi_N = (\frac{\hbar
\mathit{V}_{N} l_N}{6\pi k_B T})^{\frac{1}{2}}$, assuming that the
mean free path is governed by grain boundary scattering, $l_N \sim
10$ nm, in the corresponding gold layer. In addition, the induced
order parameter appeared to have \textit{s}-wave symmetry.

The \textit{d}-wave symmetry of the pair potential in YBCO should
also lead to anisotropy in the S-N PE for different
crystallographic orientations within the a-b plain, mainly between
anti-nodal and nodal surfaces \cite{11,12,13}. According to some
theoretical calculations, no conventional PE is expected for
junctions involving the nodal (110) YBCO surface \cite{13}. In
particular, no penetration of the pair amplitude into the N side,
which leads to the appearance of 'mini-gaps' \cite{4,5,6,7} in the
DOS, is predicted. Recent circuit theory models suggest that zero
energy Andreev bound state (ABS) channels quench the PE in the
nodal direction \cite{11,12}, since electrons can enter S and
\textit{uncorrelated} electron-hole pairs can penetrate N through
the ABS channels. However, a clear picture of the spatial
variation of the DOS in the vicinity of these junctions, in
particular the (possible) penetration of the ABS into the N layer
has not yet been established. Resolving these issues, which is the
main focus of the present paper, is important for gaining a better
understanding of both the PE involving \textit{d}-wave
superconductors and the nature of the ABS.

As mentioned above, ABS appear on the nodal surfaces of
\textit{d}-wave superconductors \cite{14,15,16,17}. These states
result from the pair breaking nature of nodal surfaces
(introducing an effective thin N layer) \cite{14}, and the
\textit{d}-wave pair potential sign inversion felt by the Andreev
reflected quasi-particles.  An intriguing question that has not
yet been examined is how the ABS propagate into a normal metal
layer that is deposited onto of the nodal S surface. One can
speculate that as long as the trajectories of the quasi-particles
are momentum and phase coherent and the relevant interfaces are
specular, ABS should reside in the N layer \cite{14}. The question
regarding how their spectral strength would evolve with the normal
layer thickness has not yet been treated, neither theoretically
nor experimentally. Specifically, a direct measurement, as
presented in this paper, showing the way in which the conductance
spectra are modified with the thickness of the N layer is still
lacking.

In this work, we studied the PE in nodal (110) epitaxial thin
films of YBCO covered with gold layers of various thicknesses. Gap
shaped tunneling spectra were sporadically measured with no clear
dependence on Au thickness. (One should bear in mind, however,
that (100) facets were scarce in these YBCO films).   At the same
time, our tunneling spectra clearly revealed, for the first time,
a penetration of the ABS into N, appearing as zero bias
conductance peaks (ZBCP) in the tunneling spectra. The penetration
depth was much shorter as compared to that observed for the order
parameter (energy gap) in (100)YBCO/Au junctions \cite{7}.
Moreover, the ZBCP did not decay exponentially, but appeared to be
nearly constant for Au thicknesses of up to $\sim 7$ nm. Above
this, the ZBCP decayed rapidly, and was not detected at all on Au
layers thicker than 10 nm. The small abundance of gaps, on one
hand, and the unique behavior of the ZBCP, on the other, point out
to an unconventional PE in S-N junctions involving nodal
\textit{d}-wave surfaces.

A total of 22 bare and gold-coated (110)YBCO samples were
measured, with Au thickness values ranging up to 30 nm. The
(110)YBCO/Au bilayers were prepared by laser ablation deposition
on (110) $SrTiO_3$ wafers. First, a 10 nm thick template layer of
YBCO was deposited at 660 $^{0}$C substrate temperature to ensure
an undisturbed (110) orientation. Subsequently, the temperature
was raised to 760 $^{0}$C, and a 50 nm thick YBCO film was
prepared, maintaining the (110) orientation. The temperature was
then lowered to 450 $^{0}$C in 50 Torr of oxygen pressure, and the
film was left for oxygen intake at this temperature for 1 h. Later
on, the gold layer was deposited \textit{in-situ} at 200 $^{0}$C,
and annealed in 250 Torr of oxygen pressure at this temperature
for about 2 h. X-ray diffraction analysis showed that the YBCO
films are oriented in the (110) direction, with less than 5\% of
other orientations. The transition temperatures of our films were
around 88 K with a relatively narrow transition width of 2 K,
implying slightly underdoped homogeneous films.

The samples were transferred from the deposition chamber in dry
atmosphere and inserted into our cryogenic homemade STM after
being exposed to ambient atmosphere for less than 10 minutes. The
STM was then cooled to 4.2 K via helium exchange gas for
measurements, performed using a Pt-Ir tip. The tunneling spectra
(dI/dV vs. V characteristics, proportional the local
quasi-particle DOS) were obtained by numerical differentiation of
I-V curves taken while momentarily disconnecting the feedback
loop.  The spectra were taken at specific well-defined locations
correlated with the surface topography. Several I-V curves were
acquired at each position to ensure data reproducibility. The
voltage and current settings (i.e. the tip-sample distance) had no
influence on the main spectral features, ruling out possible
contributions of single electron charging effects \cite{18}.

\begin{figure}
\includegraphics[width=10cm]{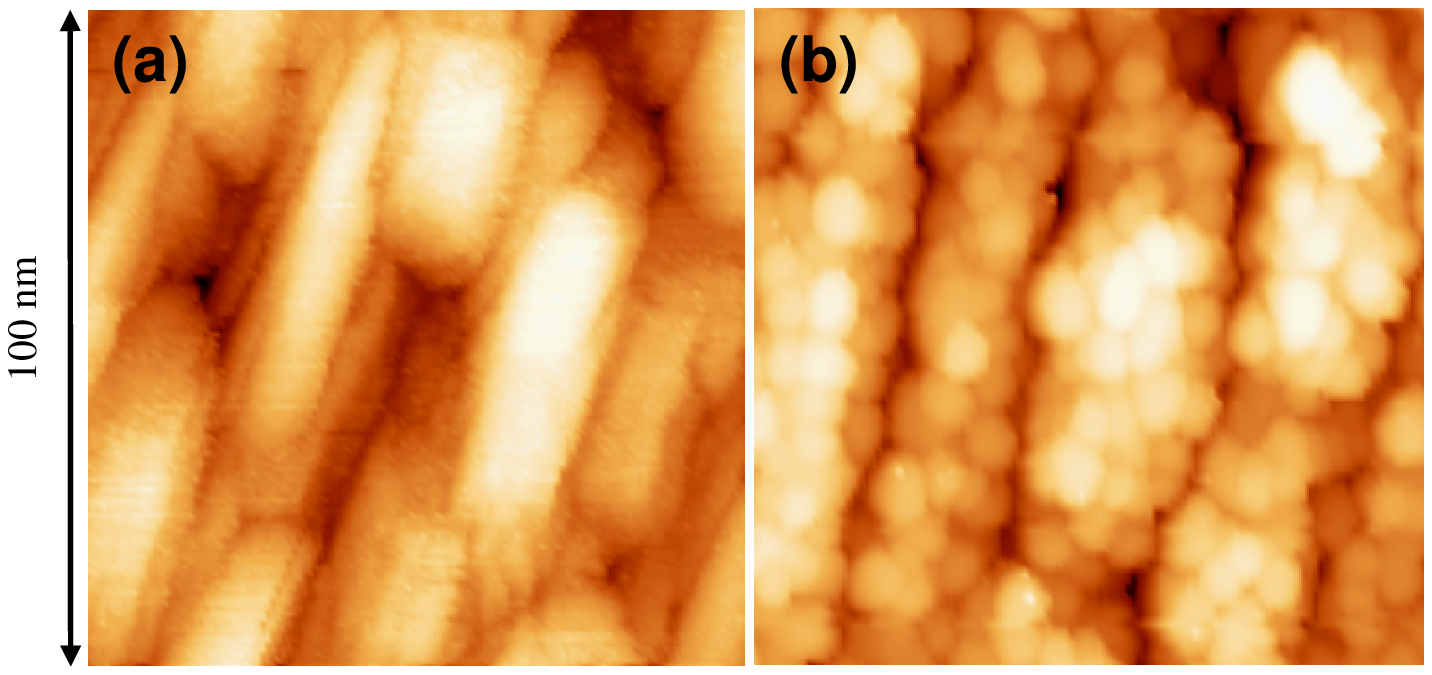}
\caption{STM topographic images of: (a) Bare (110) YBCO surface
showing elongated crystallite structures with average height of 4
nm. (b) 7 nm (nominal) gold film coating a similar YBCO sample.
The underlying YBCO structure is still visible under the Au layer
having rms roughness of 1.5 nm. }\label{Fig1}
\end{figure}

STM and AFM measurements of the bare YBCO films revealed
$\sim40\times100$ nm$^{2}$ elongated crystallite structures about
4 nm in height. The crystallites had uniform directionality,
parallel to the (110) side of the $SrTiO_3$ wafer, over areas of a
few $\mu$m$^{2}$, as portrayed in Fig.\,1(a).  This structure is
consistent with twinning of our YBCO films. The elongated
crystallite structure was clearly visible even after deposition of
the thickest gold layer (30 nm). Figure 1(b) presents an YBCO film
coated with a 7 nm thick gold layer, clearly showing both the
finer Au granularity and the elongated crystallite structure of
the underlying YBCO film. The figure also shows that the gold
layer fully covered the YBCO film and the surface morphology
revealed grains with lateral size of 10 nm and rms height
roughness of less than 1.5 nm.  We note that after annealing, good
coverage (with no apparent "pin-holes") was achieved for gold
films of average thickness larger than 3-4 nm.

\begin{figure}
\includegraphics[width=10cm]{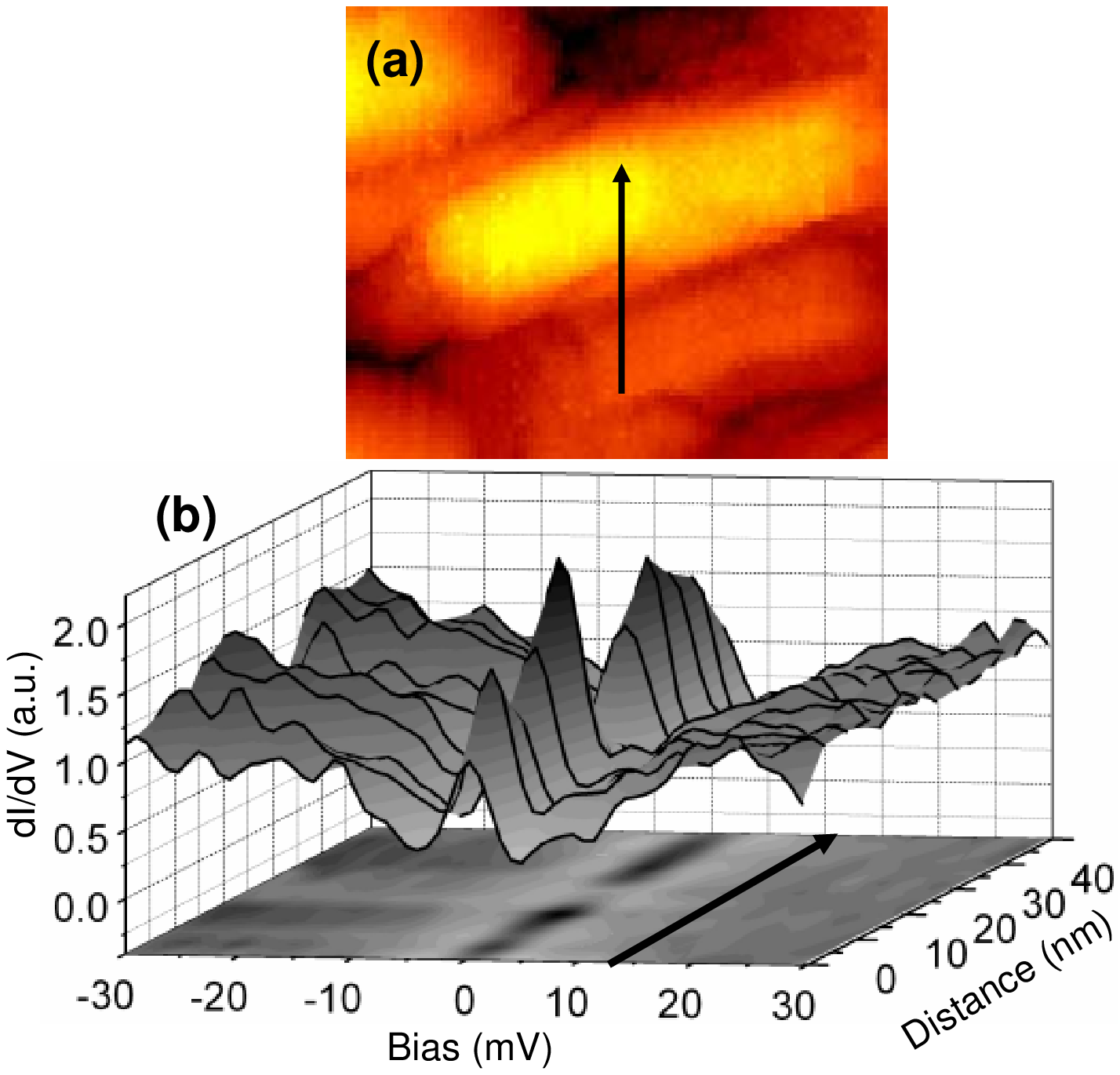}
\caption{STM measurement demonstrating the spatial evolution of
the ZBCP on the bare YBCO films. (a) $100\times 70$ nm$^{2}$
topographic image, featuring two adjacent YBCO crystallites
(possibly twins). (b) Tunneling spectra taken along the arrow
marked in (a). The ZBCP is continuous over the YBCO crystallite
but vanishes near the boundary. The projection of the spectra (in
gray scale) onto the XY plane clearly portrays the vanishing of
the ZBCP near the boundary and its nearly constant
width.}\label{Fig2}
\end{figure}

The bare samples exhibited pronounced ZBCP in the tunneling
spectra over large areas, indicative of a dominant (110) surface
orientation. In addition to the ZBCP, the spectra  also show
gap-like features at approximately 10 meV as well as an asymmetry
between the negative and positive bias, with the negative side
being steeper than the positive side. A correlated
topography-spectroscopy measurement manifesting the spatial
changes of the ZBCP on the bare YBCO films is presented in Fig. 2.
Figure 2(a) presents a topographic image of a bare YBCO film,
focusing on two adjacent crystallites. The tunneling spectra shown
in Fig. 2(b) were sequentially obtained at fixed steps along the
arrow marked in Fig. 2(a). Evidently, the ZBCP (thus, also the
ABS) exhibit spatial continuity over the YBCO crystallite but
vanish in the vicinity of the (possibly twin) boundary (and other
surface imperfections). This may be due to the formation of a thin
disordered layer near the boundaries, in which the \textit{d}-wave
order parameter is strongly reduced and in particular the ZBCP is
smeared out \cite{19,20}. Interestingly, the ZBCP appears to
recover over a length scale that is comparable to the YBCO
coherence length, $\sim2$ nm. Note that the ZBCP had a nearly
fixed width, independent of its amplitude and the point of
acquisition (near or far from boundaries and imperfections), in
agreement with previous reports \cite{17,21}.

\begin{figure}
\includegraphics[width=10cm]{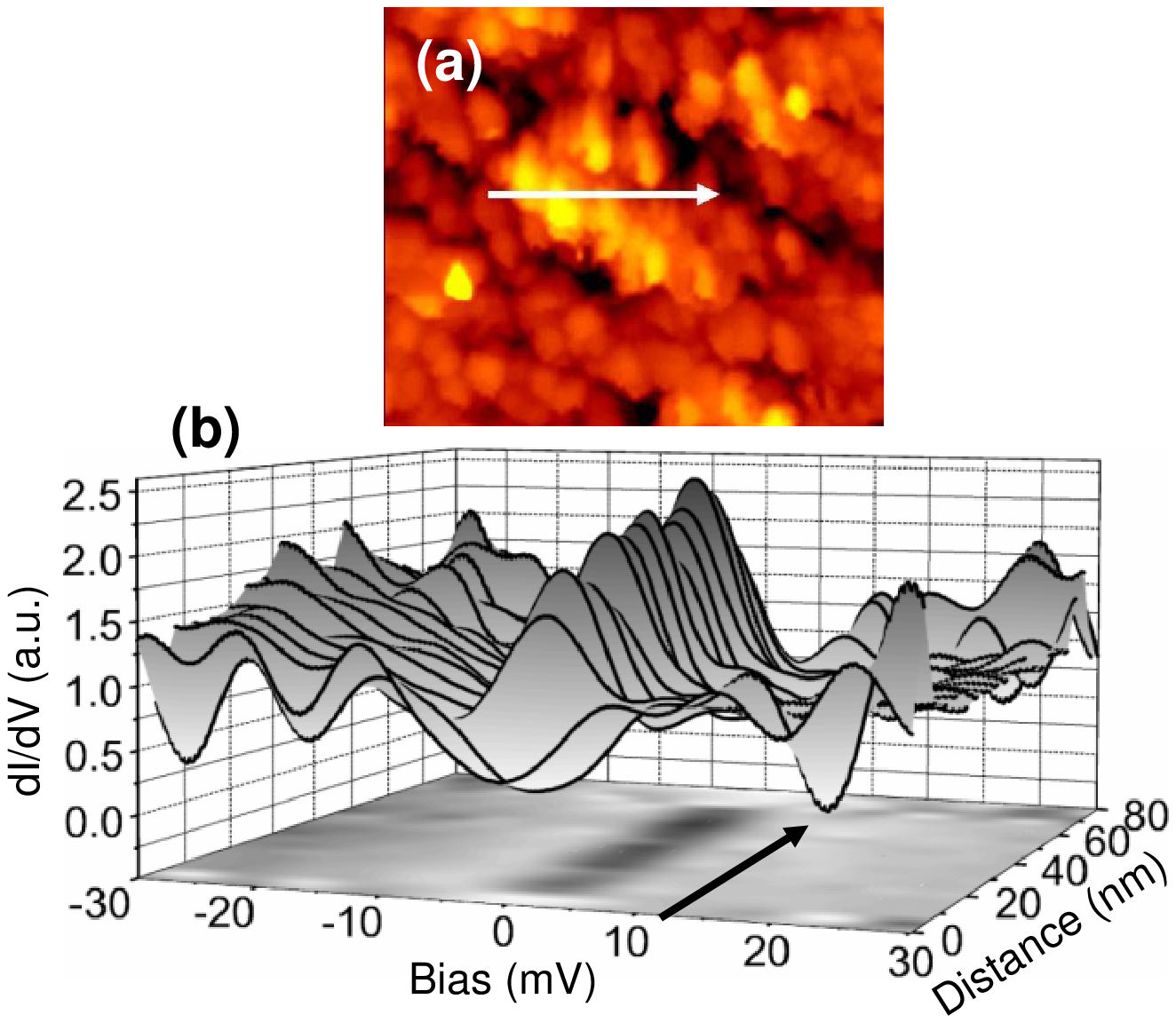}
\caption{Spatial evolution of the ZBCP in a  (110)YBCO/7nm gold
bilayer. (a)  $100\times 70$ nm$^{2}$ topographic image where the
structure of the underlying YBCO crystallite is clearly visible.
(b) Tunneling spectra measured along the white arrow spanning a
full width of a crystallite. The ZBCP is continuous over the full
width of the YBCO crystallite, with mild fluctuation, and
disappears at the edges (evident in the projection onto the XY
plane). }\label{Fig3}
\end{figure}

The ZBCP appeared not only on the bare YBCO film, but also on
samples with thin gold coating, featuring the same spatial
variation as in the bare films, as demonstrated in Fig. 3. Figure
3(a) presents a topographic image of a 7 nm YBCO/Au bilayer,
showing, again, full coverage of the Au film. The tunneling
spectra presented in Fig. 3(b) were taken along the white arrow
marked in Fig. 3(a), spanning a full cross-section of a single
crystallite. The ZBCP were observed all over the YBCO crystallite,
on and between gold grains, with mild fluctuations in height.
However it vanished at the crystallite boundaries, thus
replicating the behavior observed on the bare YBCO films.

In order to quantitatively compare the ZBCP amplitude between
samples of different gold layer thickness, we applied the
following procedure for calculating the area of the peak, which is
a measure of the density of ABS. The ZBCP were first normalized by
the normal conductance background well above the superconducting
gap, at a bias voltage of around +30 mV. Then the area between the
ZBCP and its base was integrated as illustrated in the inset of
Fig. 4.

\begin{figure}
\includegraphics[height=8 cm,angle=-90]{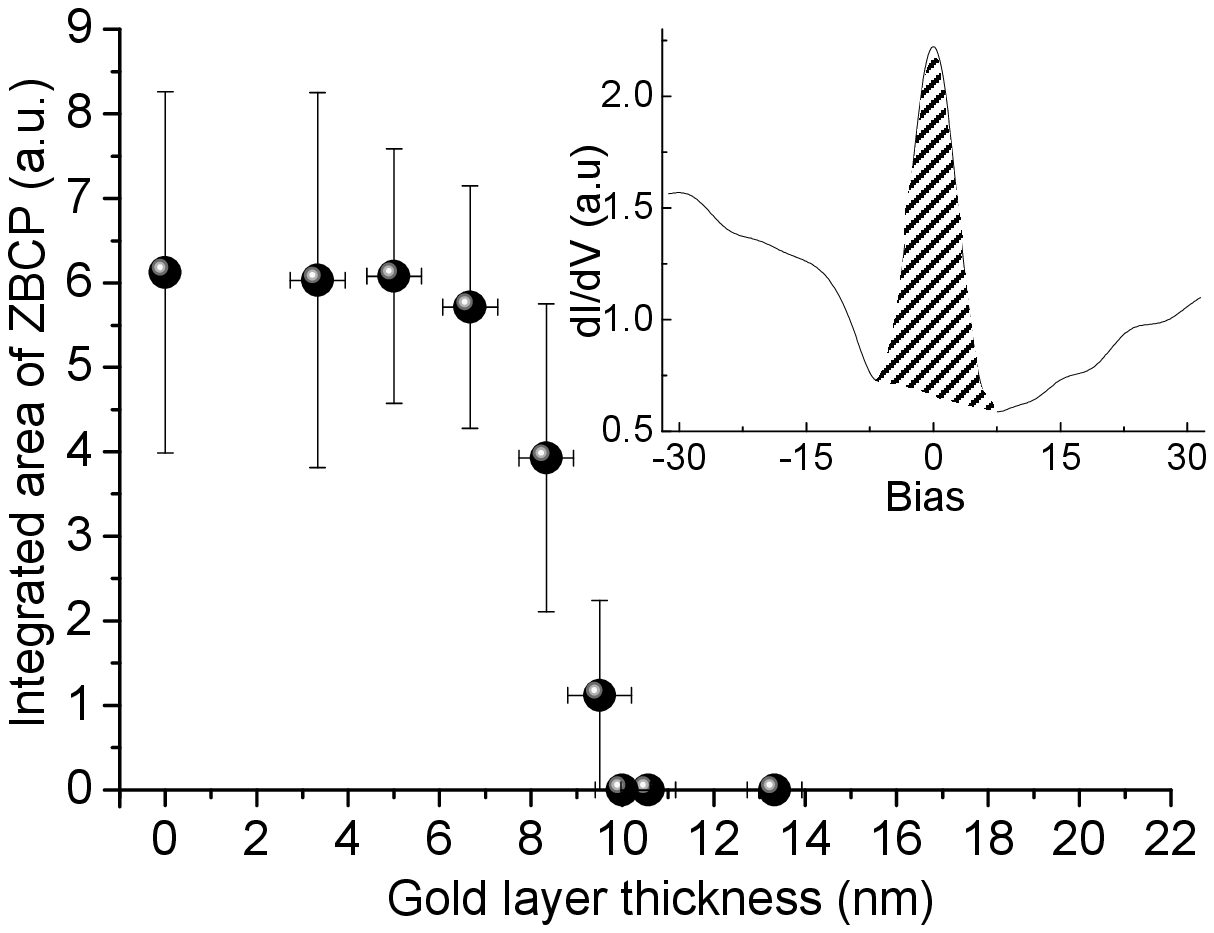}
\caption{The normalized ZBCP integrated area (proportional to the
number of ABS) as a function of nominal gold layer thickness. The
inset depicts the integrated area of the normalized peak. The
density of ABS is nearly constant up to $\sim$7nm, then decays
abruptly, and disappears above 10 nm. }\label{Fig4}
\end{figure}

Fig. 4 shows the integrated ZBCP area (proportional to the number
of ABS) as a function of the gold layer thickness, i.e., the
distance from the N-S interface. This plot depicts a behavior that
is markedly different from an exponential decay expected for
conventional proximity systems \cite{2}. An exponential decay was
observed also for the (100)YBCO/Au interface by Sharoni \textit{et
al.} \cite{7}. The magnitude of the ZBCP in Fig. 4 appears to be
rather constant for gold layers of thickness up to $\sim 7$ nm,
and decays abruptly above that.  In particular, the ABS vanished
10 nm away from the (110)YBCO/Au interface, in contrast to the
(100)YBCO/Au interface, where proximity induced energy-gaps were
still clearly observed at distances larger than 30 nm away from it
\cite{7}. We note here that the gold layers had a similar
morphology in both cases. The large vertical error bars in Fig. 4
reflect the extreme sensitivity of the ZBCP amplitude to the local
spatial variations of the disorder in the gold and the YBCO films,
and the specularity of the interface. For instance, the sample
with 9.5 nm gold thickness showed large areas free of ZBCP and
therefore the corresponding error bar extends to zero.

In order to better understand our results we recall that the mean
free path in our gold films is determined by grain-boundary
scattering and is around 10 nm \cite{7}.  This implies that as
long as the normal layer is shorter than the mean free path, it
does not have much effect on the density of ABS, as predicted by
Hu \cite{14}. Namely, the penetration range of the ZBCP is
determined by the "ballistic" mean free path, in contrast to the
penetration range of the order parameter which was found to
coincide with the thermal (or phase coherence) length, $\xi_N =
(\frac{\hbar \mathit{V}_{N} l_N}{6\pi k_B T})^{\frac{1}{2}}$. We
note in passing that Fig. 2 displays a third characteristic
length, the coherence length in YBCO, $\sim 2$ nm, which
determines the range over which the ZBCP recovers away from an
imperfection on the (110)YBCO surface.

In addition to the ZBCP, we also measured spectra exhibiting
mini-gaps. These gaps were occasionally observed only on rather
small portions of the gold-coated films. Their size did not show
any clear correlation with the thickness of the Au layer, nor with
any apparent morphological feature.  They may be associated with
(100)YBCO nano-facets (such as in Ref. \cite{7}) that are masked
in the topographic images by the Au coating. The short length
scale (smaller than  $\xi_N$ and of the order of $l_N$) over which
the ZBCP persists in the normal layer, and the sporadic nature of
the detected mini-gaps, point out to an unconventional PE in S-N
bilayers involving the nodal surface of a \textit{d}-wave
superconductor. The PE manifests itself via a unique appearance of
the ZBCP on the N surface, as demonstrated in Fig. 4 and further
discussed below.

The formation of bound states in a proximity S-N system that give
rise to the ZBCP measured on the N side requires several
conditions. First, the S-N interface should be transparent in
order to allow Andreev reflections.  This condition is satisfied
in our samples due to the \textit{in-situ} deposition and
post-annealing of the gold layer (see Ref. \cite{7}). The second
condition is having phase coherence between the electron and hole
over their trajectories in the normal layer \cite{1,14}. This
requirement is obviously maintained in our case, since the
corresponding coherence length was found  to be 30 nm \cite{7},
larger than the scale over which the ZBCPs were observed here. The
last requirement, which is related to the properties of the Au
film, is having specular reflection at the gold surface and
electron and hole momentum conservation within it \cite{14,22,23}.
The latter condition should hold for gold layers thinner than
$l_N$ (the grain size in our case), consistent with our
observation for the penetration depth of $\sim$10 nm.  However,
the unique thickness dependence of the ZBCP, exhibiting a
non-exponential decay unlike diffusive systems, still needs to be
explained. For this we need one of the above conditions to fail
abruptly.  Possibly, due to the narrow distribution of the gold
grain size (as evident from our morphological analysis of the gold
layers), there is a well-defined length scale for momentum
conservation breaking, resulting in a rapid decay of the ZBCP
amplitude. We note here that reminiscent "ballistic" effects were
also detected in conventional S-N proximity systems for gold
layers of similar granularity and thickness as ours \cite{24}.

In summary, our scanning tunneling spectroscopy of (110)YBCO/Au
bilayers features an anomalous PE. A mini-gap structure, typical
of S-N proximity systems was not induced. ABS however, clearly
penetrate the gold layer, but only over a short length determined
by the mean free path, $\sim10$ nm, which is much shorter than the
normal coherence length. The amplitude of the corresponding ZBCP
decays with gold layer thickness in a unique non-exponential way,
typical of an order parameter in diffusive proximity systems.
Rather, the amplitude is nearly constant up to a gold layer
thickness of 7 nm, and then decays abruptly, signifying a
ballistic effect.

We thank Guy Deutscher for stimulating discussions. This research
was supported in part by the Israel Science Foundation (grant No.
1565/04), the Heinrich Hertz Minerva Center for HTSC, the Karl
Stoll Chair in advanced materials, and by the Fund for the
Promotion of Research at the Technion.


\begin{references}
\bibitem {1}
E. L. {Wolf}, {\emph{Principles of Electron Tunneling
Spectroscopy}} (Oxford University Press, New York, 1985).

\bibitem {2}
G. {Deutscher} and P. G. {De Gennes}, {\emph{Superconductivity}}
(Marcel Dekker, Inc., New York, 1969).

\bibitem {3}
Y. {Levi}, \text{et al.}, Phys. Rev. B \textbf{58}, 15128 (1998).

\bibitem {4}
N. {Moussy}, H. {Courtois}, and B. {Pannetier}, Europhys. Lett.
\textbf{55}, 861 (2001).

\bibitem {5}
W. {Belzig}, C. {Bruder}, and G. {Schon}, Phys. Rev. B
\textbf{54}, 9443 (1996).

\bibitem {6}
S. {Gueron}, \text{et al.}, Phys. Rev. Lett. \textbf{77}, 3025
(1996).

\bibitem {7}
A. {Sharoni}, \text{et al.}, Phys. Rev. Lett. \textbf{92}, 017003
(2004).

\bibitem {8}
M. {Lee},  \text{et al.}, Appl. Phys. Lett \textbf{57}, 1152
(1990).

\bibitem {9}
M. A. M. {Gijs}, \text{et al.}, Appl. Phys. Lett \textbf{57}, 2600
(1990).

\bibitem {10}
H. Z. {Durusoy}, \text{et al.}, Physica C \textbf{266}, 253
(1996).

\bibitem {11}
Y. {Tanaka}, \text{et al.}, Phys. Rev. B \textbf{69}, 144519
(2004).

\bibitem {12}
Y. {Tanaka}, Y. V. {Nazarov}, and S. {Kashiwaya}, Phys. Rev. Lett.
\textbf{90}, 167003 (2003).

\bibitem {13}
Y. {Ohashi}, J.  Phys. Soc. Jpn. \textbf{65}, 823 (1996).

\bibitem {14}
C. R. {Hu}, Phys. Rev. Lett. \textbf{72}, 1526 (1994).

\bibitem {15}
Y. {Tanaka} and S. {Kashiwaya}, Phys. Rev. Lett. \textbf{74}, 3451
(1995).

\bibitem {16}
A. {Sharoni}, G. {Koren}, and O. {Millo}, Europhys. Lett.
\textbf{54}, 675 (2001).

\bibitem {17}
L. {Alff},  \text{et al.}, Phys. Rev. B \textbf{55}, R14757
(1997).

\bibitem {18}
E. {Bar-Sadeh}, \text{et al.},  Phys. Rev. B \textbf{53}, 3482
(1996).

\bibitem {19}
J. X. {Zhu},  \text{et al.}, Phys. Rev. Lett. \textbf{91}, 057004
(2003).

\bibitem {20}
A. A. {Golubov} and M. Y. {Kupriyanov}, Superlattices and
Microstructures \textbf{25}, 949 (1999).

\bibitem {21}
M. B. {Walker} and P. {Pairor}, Phys. Rev. B \textbf{60}, 10395
(1999).

\bibitem {22}
M. {Fogelstrom}, D. {Rainer}, and J. A. {Sauls}, Phys. Rev. Lett.
\textbf{79}, 281 (1997).

\bibitem {23}
Y. {Asano} and Y. {Tanaka}, Phys. Rev. B \textbf{65}, 064522
(2002).

\bibitem {24}
A. K. {Gupta}, \text{et al.}, Phys. Rev. B \textbf{69}, 104514
(2004).


\end{references}
\end{document}